\documentstyle[twocolumn,aps,floats,amssymb,epsfig]{revtex}

\begin{document}

\def\d{{\rm d}}
\def\e{{\rm e}}
\def\O{{\rm O}}
\def\half{\mbox{$\frac12$}}
\def\eref#1{(\protect\ref{#1})}
\def\etal{{\it{}et~al.}}

\draft
\tolerance = 10000

\setcounter{topnumber}{2}
\renewcommand{\topfraction}{0.9}
\renewcommand{\textfraction}{0.1}
\renewcommand{\floatpagefraction}{0.5}

\twocolumn[\hsize\textwidth\columnwidth\hsize\csname @twocolumnfalse\endcsname

\title{Renormalization group analysis of the small-world network model}
\author{M. E. J. Newman and D. J. Watts}
\address{Santa Fe Institute, 1399 Hyde Park Road, Santa Fe, NM 87501}
\maketitle

\begin{abstract}
We study the small-world network model, which mimics the transition between
regular-lattice and random-lattice behavior in social networks of
increasing size.  We contend that the model displays a normal continuous
phase transition with a divergent correlation length as the degree of
randomness tends to zero.  We propose a real-space renormalization group
transformation for the model and demonstrate that the transformation is
exact in the limit of large system size.  We use this result to calculate
the exact value of the single critical exponent for the system, and to
derive the scaling form for the average number of ``degrees of separation''
between two nodes on the network as a function of the three independent
variables.  We confirm our results by extensive numerical simulation.
\end{abstract}

\vspace{0.5cm}

]

Folk wisdom holds that there are ``six degrees of separation'' between any
two human beings on the planet---i.e.,~a path of no more than six
acquaintances linking any person to any other.  While the exact number six
may not be a very reliable estimate, it does appear that for most social
networks quite a short chain is needed to connect even the most distant of
the network's members\cite{milgram67}, an observation which has important
consequences for issues such as the spread of disease\cite{SS88},
oscillator synchrony\cite{kuramoto84}, and genetic regulatory
networks\cite{kauffman69}.  At first sight this does not seem too
surprising a result; random networks have average vertex--vertex distances
which increase as the logarithm of the number of vertices and which can
therefore be small even in very large networks\cite{bollobas85}.  However,
real social networks are far from random, possessing well-defined locales
in which the probability of connection is high and very low probability of
connection between two vertices chosen at random.  Watts and
Strogatz\cite{WS98} have recently proposed a model of the ``small world''
which reconciles these observations.  Their model does indeed possess
well-defined locales, with vertices falling on a regular lattice, but in
addition there is a fixed density of random ``shortcuts'' on the lattice
which can link distant vertices.  Their principal finding is that only a
very small density of such shortcuts is necessary to produce vertex--vertex
distances comparable to those found on a random lattice.

In this paper we study the model of Watts and Strogatz using the techniques
of statistical physics, and show that it possesses a continuous phase
transition in the limit where the density of shortcuts tends to zero.  We
investigate this transition using a renormalization group (RG) method and
calculate the scaling forms and the single critical exponent describing the
behavior of the model in the critical region.

\begin{figure}
\begin{center}
\psfig{figure=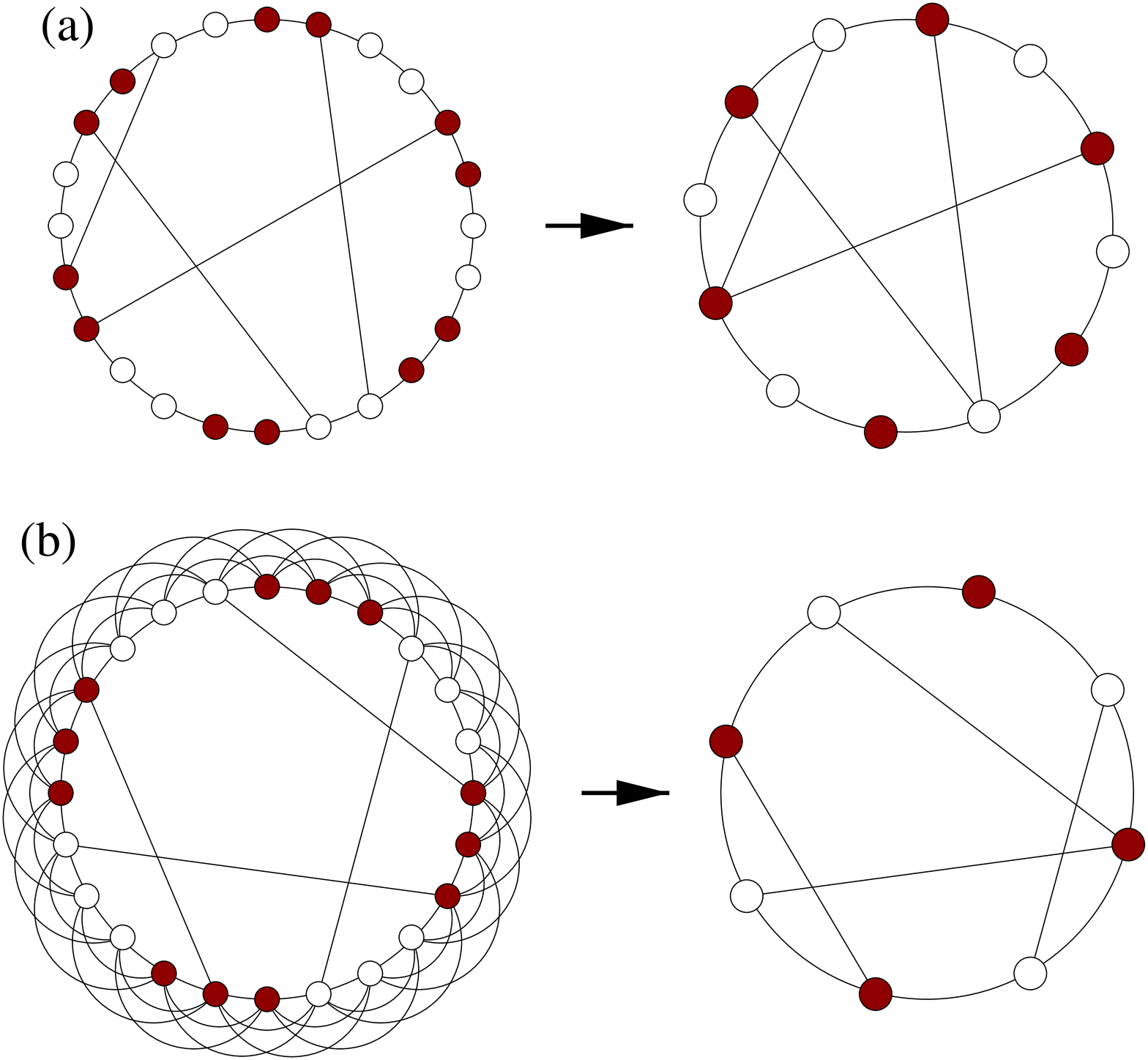,width=3.25in}
\end{center}
\caption{The RG transformations used in the calculations described in the
  text: (a)~the transformation used for the $k=1$ system; (b)~the
  transformation used for $k>1$, illustrated in this case for $k=3$.  The
  coloring of the sites indicates how they are grouped under the
  transformations.}
\label{rg}
\end{figure}

Previous studies have concentrated on the one-dimensional version of the
small-world model, and we will start with this version too, although we
will later generalize our results to higher dimensions.  In one dimension
the model is defined on a lattice with $L$ sites and periodic boundary
conditions (the lattice is a ring).  Initially each site is connected to
all of its neighbors up to some fixed range $k$ to make a network with
average coordination number $z=2k$\cite{note3}.  Randomness is then
introduced by independently rewiring each of the $kL$ connections with
probability $p$.  ``Rewiring'' in this context means moving one end of the
connection to a new, randomly chosen site.  The behavior of the network
thus depends on three independent parameters: $L$, $k$ and $p$.  In this
paper we will study a slight variation on the model in which shortcuts are
added between randomly chosen pairs of sites, but no connections are
removed from the regular lattice.  For sufficiently small $p$ and large $L$
this makes no difference to the mean separation between vertices of the
network for $k\ge2$.  For $k=1$ it does make a difference, since the
original small-world model is poorly defined in this case---there is a
finite probability of a part of the lattice becoming disconnected from the
rest and therefore making an infinite contribution to the average distance
between vertices, and this makes the distance averaged over all networks
for a given value of $p$ also infinite.  Our variation does not suffer from
this problem and this makes the analysis significantly simpler.  In
Fig.~\ref{rg} we show some examples of small-world networks.

We consider the behavior of the model for low density $p$ of shortcuts.
The fundamental observable quantity that we measure is the shortest
distance between a pair of vertices on the network, averaged both over all
pairs on the network and over all possible realizations of the randomness.
This quantity, which we denote $\ell$, has two regimes of behavior.  For
systems small enough that there is much less than one shortcut on the
lattice on average, $\ell$ is dominated by the connections of the regular
lattice and can be expected to increase linearly with system size $L$.  As
the lattice becomes larger with $p$ held fixed, the average number of
shortcuts will eventually become greater than one and $\ell$ will start to
scale as $\log L$.  The transition between these two regimes takes place at
some intermediate system size $L=\xi$, and from the arguments above we
would expect $\xi$ to take a value such that the number of shortcuts $p k
\xi\simeq1$.  In other words we expect $\xi$ to diverge in the limit of
small $p$ as $\xi\sim p^{-1}$.  The quantity $\xi$ plays a role similar to
the correlation length in an interacting system in conventional statistical
physics, and its divergence leaves the small-world model with no
characteristic length scale other than the fundamental lattice spacing.
Thus the model possesses a continuous phase transition at $p=0$, and, as we
will see, this gives rise to specific finite-size scaling behavior in the
region close to the transition.  Note that the transition is a one-sided
one, since $p$ can never take a value less than zero.  In this respect the
transition is similar to transitions seen in other one-dimensional systems
such as 1D bond or site percolation, or the 1D Ising model.

Barth\'el\'emy and Amaral\cite{BA99} have suggested that the arguments
above, although correct in outline, are not correct in detail.  They
contend that the length-scale $\xi$ diverges as
\begin{equation}
\xi \sim p^{-\tau}
\label{defstau}
\end{equation}
with $\tau$ different from the value of~1 given by the scaling argument.
On the basis of numerical results, they conjecture that $\tau=\frac23$.
Barret\cite{barret99}, on the other hand, has given a simple physical
argument which directly contradicts this, indicating that $\tau$ should be
greater than or equal to~1.  Amongst other things, we demonstrate in this
paper that in fact $\tau$ is exactly~1 for all values of $k$.

Let us first consider the small-world model for the simplest case $k=1$.
As discussed above, the average distance $\ell$ scales linearly with $L$
for $L\ll\xi$ and logarithmically for $L\gg\xi$.  If $\xi$ is the only
non-trivial length-scale in the problem and is much larger than one
(i.e.,~we are close to the phase transition), this implies that $\ell$
should obey a finite-size scaling law of the form
\begin{equation}
\ell = L f(L/\xi),
\label{scaling1}
\end{equation}
where $f(x)$ is a universal scaling function with the limiting forms
\begin{equation}
f(x) \sim \Bigl\lbrace \begin{array}{ll}
            \mbox{constant}\qquad & \mbox{for $x\ll1$}\\
            (\log x)/x & \mbox{for $x\gg1$.}
          \end{array}
\label{limits}
\end{equation}
In fact, it is easy to show that the limiting value of $f(x)$ as $x\to0$ is
$\frac14$.  A scaling law similar to this has been proposed previously by
Barth\'el\'emy and Amaral\cite{BA99} for the small-world model, although
curiously they suggested that scaling of this type was evidence for the
{\em absence\/} of a phase transition in the model, whereas we regard it as
the appropriate form for $\ell$ in the presence of one\cite{note2}.

We now assume that, in the critical region, $\xi$ takes the
form~\eref{defstau}, and that we do not know the value of the exponent
$\tau$.  Then we can rewrite Eq.~\eref{limits} in the form
\begin{equation}
\ell = L f(p^\tau L),
\label{scaling2}
\end{equation}
where we have absorbed a multiplicative constant into the argument of
$f(x)$, but otherwise it is the same scaling function as before, with the
same limits, Eq.~\eref{limits}.

Now consider the real-space RG transformation on the $k=1$ small-world
model in which we block sites in adjacent pairs to create a one-dimensional
lattice of a half as many sites.  (We assume that the lattice size $L$ is
even.  In fact the transformation works fine if we block in groups of any
size which divides $L$.)  Two vertices are connected on the renormalized
lattice if either of the original vertices in one was connected to either
of the original vertices in the other.  This includes shortcut connections.
The transformation is illustrated in Fig.~\ref{rg}a for a lattice of size
$L=24$.

The number of shortcuts on the lattice is conserved under the
transformation, so the fundamental parameters $L$ and $p$ renormalize
according to
\begin{equation}
L' = \half L,\qquad p' = 2p.
\label{rg1}
\end{equation}
The transformation generates all possible configurations of shortcuts on
the renormalized lattice with the correct probability, as we can easily see
since the probability of finding a shortcut between any two sites $i$ and
$j$ is uniform, independent of $i$ and $j$ both before and after
renormalization.  The geometry of the shortest path between any two points
is unchanged under our transformation.  However, the length of the path is,
on average, halved along those portions of the path which run around the
perimeter of the ring, and remains the same along the shortcuts.  For large
$L$ and small $p$, the portion of the length along the shortcuts tends to
zero and so can be neglected.  Thus
\begin{equation}
\ell' = \half\ell
\label{rg2}
\end{equation}
in this limit.

Eqs.~\eref{rg1} and~\eref{rg2} constitute the RG equations for this system
and are exact for $n\gg1$ and $p\ll1$.  Substituting into
Eq.~\eref{scaling2} we then find that
\begin{equation}
\tau = {\log(L/L')\over\log(p'/p)} = 1.
\end{equation}

Now we turn to the case of $k>1$.  To treat this case we define a slightly
different RG transformation: we group adjacent sites in groups of $k$, with
connections assigned using the same rule as before.  The transformation is
illustrated in Fig.~\ref{rg}b for a lattice of size $L=24$ with $k=3$.
Again the number of shortcuts in the network is preserved under the
transformation, which gives the following renormalization equations for the
parameters:
\begin{equation}
L' = L/k,\qquad p' = k^2 p,\qquad k' = 1,\qquad \ell' = \ell.
\label{rg3}
\end{equation}
Note that, in the limit of large $L$ and small $p$, the mean distance
$\ell$ is not affected at all; the number of vertices along the path
joining two distant sites is reduced by a factor $k$, but the number of
vertices that can be traversed in one step is reduced by the same factor,
and the two cancel out.  For the same reasons as before, this
transformation is exact in the limit of large $L$ and small $p$.

We can use this second transformation to turn any network with $k>1$ into a
corresponding network with $k=1$, which we can then treat using the
arguments given before.  Thus, we conclude, the exponent $\tau=1$ for all
values of $k$ and, substituting from Eq.~\eref{rg3} into
Eq.~\eref{scaling2}, the general small-world network must satisfy the
scaling form
\begin{equation}
\ell = {L\over k} f(pkL).
\label{scaling3}
\end{equation}
This form should be correct for $L'\gg1$ and $p'\ll1$, which implies that
$L/k\gg1$ and $k^2 p\ll1$.  The first of these conditions is trivial---it
merely precludes inaccuracies of $\pm k$ in the estimate of $\ell$ because
positions on the lattice are rounded off to the nearest multiple of $k$ by
the RG transformation.  The second condition is interesting however; it is
necessary to ensure that the average distance traveled along shortcuts in
the network is small compared to the distance traveled around the perimeter
of the ring.  This condition tells us when we are moving out of the scaling
regime close to the transition, which is governed by~\eref{scaling3}, into
the regime of the true random network, for which~\eref{scaling3} is badly
violated and $\ell$ is known to scale as $\log L/\log k$\cite{bollobas85}.
It implies that we need to work with values of $p$ which decrease as
$k^{-2}$ with increasing $k$ if we wish to see clean scaling behavior, or
conversely, that true random-network behavior should be visible in networks
with values of $p\simeq k^{-2}$ or greater.

\begin{figure}
\begin{center}
\psfig{figure=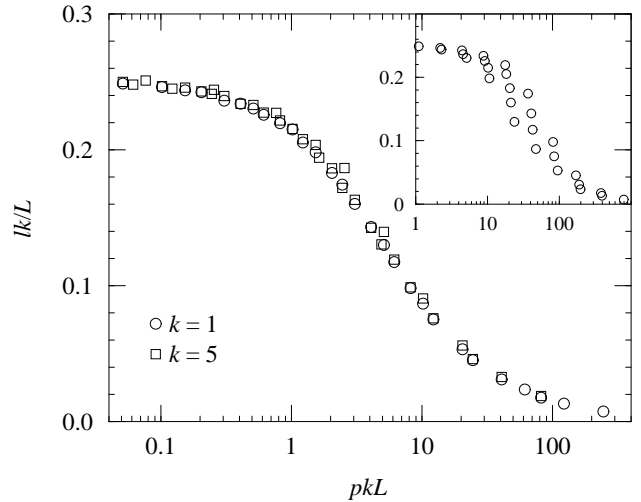,width=3.25in}
\end{center}
\caption{Collapse of numerical data for $\ell$ according to
Eq.~\eref{scaling3} as described in the text.  Error bars are in all cases
smaller than the data points.  Note that the horizontal axis is
logarithmic.  Inset: the collapse is noticeably poorer for $\tau=\frac23$.}
\label{collapse}
\end{figure}

We have tested our predictions by extensive numerical simulation of the
small-world model.  We have calculated exhaustively the minimum distance
between all pairs of points on a variety of networks and averaged the
results to find $\ell$.  We have done this for $k=1$ (coordination number
$z=2$) for systems of size $L$ equal to a power of two from $128$ up to
$8192$ and $p=1\times10^{-4}$ up to $3\times10^{-2}$, and for $k=5$
($z=10$) with $L=512\ldots 32\,768$ and $p=3\times10^{-6} \ldots
1\times10^{-3}$.  Each calculation was averaged over 1000 realizations of
the randomness.  In Fig.~\ref{collapse} we show our results plotted as the
values of $\ell k/L$ against $pkL$.  Eq.~\eref{scaling3} predicts that when
plotted in this way, the results should collapse onto a single curve and,
as the figure shows, they do indeed do this to a reasonable approximation.

As mentioned above, Barth\'el\'emy and Amaral\cite{BA99} also performed
numerical simulations of the small-world model and extracted a value of
$\tau=\frac23$ for the critical exponent.  In the inset of
Fig.~\ref{collapse} we show our simulation results for $k=1$ plotted
according to Eq.~\eref{scaling2} using this value for $\tau$.  As the
figure shows, the data collapse is significantly poorer in this case than
for $\tau=1$.  It is interesting to ask then how Barth\'el\'emy and Amaral
arrived at their result.  It seems likely that the problem arises from
looking at systems that are too small to show the true scaling behavior.
In our calculations, we find good scaling for $L/k\gtrsim60$.
Barth\'el\'emy and Amaral examined networks with $k=5$, 10 and 15 ($z=10$,
20, 30) so we should expect to find good scaling behavior for values of $L$
larger than about 600.  However, the systems studied by Barth\'el\'emy and
Amaral ranged in size from about $L=50$ to about $500$ in most cases, and
in no case exceeded $L=1000$.  Their calculations therefore had either no
overlap with the scaling regime, or only a small overlap, and so we would
not expect to find behavior typical of the true value of $\tau$ in their
results.

\begin{figure}
\begin{center}
\psfig{figure=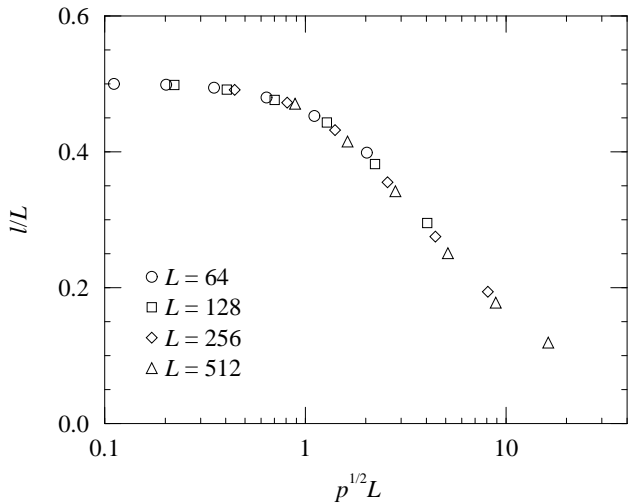,width=3.25in}
\end{center}
\caption{Collapse of numerical data for networks based on the square lattice
  in two dimensions, as described in the text.  Error bars are in all cases
  smaller than the data points.}
\label{higher}
\end{figure}

It is possible to generalize the calculations presented here to small-world
networks built on lattices of dimension $d$ greater than one.  For
simplicity we consider first the case $k=1$.  If we construct a square or
(hyper)cubic lattice in $d$ dimensions with linear dimension $L$,
connections between nearest neighbor vertices, and shortcuts added with a
rewiring probability of $p$, then as before the average vertex--vertex
distance scales linearly with $L$ for small $L$, logarithmically for large
$L$, and the length-scale $\xi$ of the transition diverges according to
Eq.~\eref{defstau} for small $p$.  Thus the scaling form~\eref{scaling2}
applies for general $d$ also.  The appropriate generalization of our RG
transformation involves grouping sites in square or cubic blocks of side~2,
and the quantities $L$, $p$ and $\ell$ then renormalize according to
\begin{equation}
L' = \half L,\qquad p' = 2^d p,\qquad \ell' = \half\ell.
\end{equation}
Thus
\begin{equation}
\tau = {\log(L/L')\over\log(p'/p)} = {1\over d}.
\end{equation}
As an example, we show in Fig.~\ref{higher} numerical results for the $d=2$
case, for $L$ equal to a power of two from $64$ up to $512$ (i.e.,~a little
over a quarter of a million vertices for the largest networks simulated)
and six different values of $p$ for each system size from
$p=3\times10^{-6}$ up to $1\times10^{-3}$.  The results are plotted
according to Eq.~\eref{scaling2} with $\tau=\half$ and, as the figure
shows, they again collapse nicely onto a single curve.

A number of generalizations are possible for $k>1$.  Perhaps the simplest
is to add connections along the principal axes of the lattice between all
vertices whose separation is $k$ or less.  This produces a graph with
average coordination number $z=2dk$.  By blocking vertices in square or
cubic blocks of edge $k$, we can then transform this system into one with
$k=1$.  The appropriate generalization of the RG equations~\eref{rg3} is
then
\begin{equation}
L' = L/k,\quad p' = k^{d+1} p,\quad
k' = 1,\quad \ell' = \ell,
\label{rg4}
\end{equation}
which gives $\tau=1/d$ for all $k$ and a scaling form of
\begin{equation}
\ell = {L\over k} f\bigl((pk)^{1/d} L\bigr).
\end{equation}
Alternatively, we could redefine our scaling function $f(x)$ so that $\ell
k/L$ is given as a function of $pkL^d$.  Writing it in this form makes it
clear that the number of vertices in the network at the transition from
large- to small-world behavior diverges as $(pk)^{-1}$ in any number of
dimensions.

Another possible generalization to $k>1$ is to add connections between all
sites within square or cubic regions of side $2k$.  This gives a different
dependence on $k$ in the scaling relation, but $\tau$ still equal to $1/d$.

To conclude, we have studied the small-world network model of Watts and
Strogatz using an asymptotically exact real-space renormalization group
method.  We find that in all dimensions $d$ the model undergoes a
continuous phase transition as the density $p$ of shortcuts tends to zero
and that the characteristic length $\xi$ diverges according to $\xi\sim
p^{-\tau}$ with $\tau=1/d$ for all values of the connection range $k$.  We
have also deduced the general finite-size scaling law which describes the
variation of the mean vertex--vertex separation as a function of $p$, $k$
and the system size $L$.  We have performed extensive numerical
calculations which confirm our analytic results.


\begin{references}
%
\bibitem{milgram67}
  S. Milgram, ``The small world problem'', {\it Psychol. Today\/} {\bf2},
  60--67 (1967).
%
\bibitem{SS88}
  L. Sattenspiel and C. P. Simon, ``The spread and persistence of
  infectious diseases in structured populations'', {\it Mathematical
    Biosciences\/} {\bf90}, 367-383 (1988).
%
\bibitem{kuramoto84}
  Y. Kuramoto, {\it Chemical Oscillations, Waves and Turbulence,} Springer,
  Berlin (1984).
%
\bibitem{kauffman69}
  S. A. Kauffman, ``Metabolic stability and and epigenesis in randomly
  constructed genetic nets'', {\it J. Theor. Biol.} {\bf22}, 437--467.
%
\bibitem{bollobas85}
B. Bollob\'as, {\it Random Graphs,} Academic Press, New York (1985).
%
\bibitem{WS98}
  D. J. Watts and S. H. Strogatz, ``Collective dynamics of small-world
  networks'', Nature {\bf393}, 440--442 (1998).
%
\bibitem{note3}
  In Ref.~\onlinecite{WS98} $k$ is defined to be equal to the coordination
  number $z$.  Here we use $k=\half z$ to avoid unnecessary factors of~2 in
  our equations.
%
\bibitem{BA99}
  M. Barth\'el\'emy and L. A. N. Amaral, ``Small-world networks: Evidence
  for a crossover picture'', {\tt cond-mat/9903108}.
%
\bibitem{barret99}
  A. Barrat, ``Comment on `Small-world networks: Evidence for a crossover
  picture','' {\tt cond-mat/9903323}.
%
\bibitem{note2}
  Ref.~\onlinecite{BA99} is a little confusing in this respect.  It appears
  the authors may have been referring to the possibility that the system
  shows a phase transition as the size $L$ of the system is varied.  This
  however would not be a sensible suggestion, since it is well-known that
  systems of finite size do not show sharp phase transitions.  The only
  sensible scenario is a phase transition with varying shortcut probability
  $p$, which the model does indeed seem to show.
%
\end{references}
\end{document}